\documentclass[master]{ws-rv9x6} % 'master' to combine all chapters
\usepackage{ws-rv-har}    % author-date citation/references %%% WS 04-Dec-17
\usepackage{ws-rv-thm}    % comment this line when `amsthm / theorem / ntheorem` package is used
\usepackage{subfigure}    % required only when side-by-side / subfigures are used
\usepackage{ws-index}     % required only when multiple indexes are used
\usepackage{enumitem}
\usepackage[percent]{overpic}
\usepackage{xcolor}
\makeindex
\newindex{aindx}{adx}{and}{Author Index}       % author index
\renewindex{default}{idx}{ind}{Subject Index}  % subject index

\begin{document}		
\newcommand{\ltsima}{$\; \buildrel < \over \sim \;$}
\newcommand{\lsim}{\lower.5ex\hbox{\ltsima}}
\newcommand{\gtsima}{$\; \buildrel > \over \sim \;$}
\newcommand{\gsim}{\lower.5ex\hbox{\gtsima}}
\newcommand{\bra}{\langle}
\newcommand{\ket}{\rangle}
\newcommand{\lprime}{\ell^\prime}
\newcommand{\lpp}{\ell^{\prime\prime}}
\newcommand{\mprime}{m^\prime}
\newcommand{\mpp}{m^{\prime\prime}}
\newcommand{\ci}{\mathrm{i}}
\newcommand{\dd}{\mathrm{d}}
\newcommand{\veck}{\mathbf{k}}
\newcommand{\vecx}{\mathbf{x}}
\newcommand{\vecr}{\mathbf{r}}
\newcommand{\vecv}{\mathbf{\upsilon}}
\newcommand{\vecw}{\mathbf{\omega}}
\newcommand{\vecj}{\mathbf{j}}
\newcommand{\vecq}{\mathbf{q}}
\newcommand{\vecl}{\mathbf{l}}
\newcommand{\vecn}{\mathbf{n}}
\newcommand{\lm}{\ell m}
\newcommand{\that}{\hat{\theta}}
\newcommand{\thatp}{\that^\prime}
\newcommand{\chip}{\chi^\prime}
\newcommand{\hs}{\hspace{1mm}}
\newcommand{\nar}{New Astronomy Reviews}
\def\gsim{~\rlap{$>$}{\lower 1.0ex\hbox{$\sim$}}}
\def\lsim{~\rlap{$<$}{\lower 1.0ex\hbox{$\sim$}}}
\def\Msun {\,\mathrm{M}_\odot}
\def\Jcrit {J_\mathrm{crit}}
\newcommand{\rsun}{R_{\odot}}
\newcommand{\mbh}{M_{\rm BH}}
\newcommand{\Msunyr}{M_\odot~{\rm yr}^{-1}}
\newcommand{\mdot}{\dot{M}_*}
\newcommand{\ledd}{L_{\rm Edd}}
\newcommand{\cmc}{{\rm cm}^{-3}}
\def\gsim{~\rlap{$>$}{\lower 1.0ex\hbox{$\sim$}}}
\def\lsim{~\rlap{$<$}{\lower 1.0ex\hbox{$\sim$}}}
\def\Msun {\,\mathrm{M}_\odot}
\def\Jcrit {J_\mathrm{crit}}

\def\simgreat{\lower2pt\hbox{$\buildrel {\scriptstyle >}
   \over {\scriptstyle\sim}$}}
\def\simless{\lower2pt\hbox{$\buildrel {\scriptstyle <}
   \over {\scriptstyle\sim}$}}
\def\msobh{M_\bullet^{\rm sBH}}
\def\zodot{\,{\rm Z}_\odot}
\newcommand{\lambdabar}{\mbox{\makebox[-0.5ex][l]{$\lambda$} \raisebox{0.7ex}[0pt][0pt]{--}}}

\def\na{NewA}%
          % New~Astronomy
\def\aj{AJ}%
          % Astronomical Journal
\def\araa{ARA\&A}%
          % Annual Review of Astron and Astrophys
\def\apj{ApJ}%
          % Astrophysical Journal
\def\apjl{ApJ}%
          % Astrophysical Journal, Letters
\def\jcap{JCAP}

\def\pasa{PASA}

\def\apjs{ApJS}%
          % Astrophysical Journal, Supplement
\def\ao{Appl.~Opt.}%
          % Applied Optics
\def\apss{Ap\&SS}%
          % Astrophysics and Space Science
\def\aap{A\&A}%
          % Astronomy and Astrophysics
\def\aapr{A\&A~Rev.}%
          % Astronomy and Astrophysics Reviews
\def\aaps{A\&AS}%
          % Astronomy and Astrophysics, Supplement
\def\azh{AZh}%
          % Astronomicheskii Zhurnal
\def\baas{BAAS}%
          % Bulletin of the AAS
\def\jrasc{JRASC}%
          % Journal of the RAS of Canada
\def\memras{MmRAS}%
          % Memoirs of the RAS
\def\mnras{MNRAS}%
          % Monthly Notices of the RAS
\def\pra{Phys.~Rev.~A}%
          % Physical Review A: General Physics
\def\prb{Phys.~Rev.~B}%
          % Physical Review B: Solid State
\def\prc{Phys.~Rev.~C}%
          % Physical Review C
\def\prd{Phys.~Rev.~D}%
          % Physical Review D
\def\pre{Phys.~Rev.~E}%
          % Physical Review E
\def\prl{Phys.~Rev.~Lett.}%
\def\pasp{PASP}%
          % Publications of the ASP
\def\pasj{PASJ}%
          % Publications of the ASJ
\def\qjras{QJRAS}%
          % Quarterly Journal of the RAS
\def\skytel{S\&T}%
          % Sky and Telescope
\def\solphys{Sol.~Phys.}%
          % Solar Physics

          % Solar Physics
\def\sovast{Soviet~Ast.}%
          % Soviet Astronomy
\def\ssr{Space~Sci.~Rev.}%
          % Space Science Reviews
\def\zap{ZAp}%
          % Zeitschrift fuer Astrophysik
\def\nat{Nature}%
          % Nature
\def\iaucirc{IAU~Circ.}%
          % IAU Cirulars
\def\aplett{Astrophys.~Lett.}%
          % Astrophysics Letters
\def\apspr{Astrophys.~Space~Phys.~Res.}%
          % Astrophysics Space Physics Research
\def\bain{Bull.~Astron.~Inst.~Netherlands}%
          % Bulletin Astronomical Institute of the Netherlands
\def\fcp{Fund.~Cosmic~Phys.}%
          % Fundamental Cosmic Physics
\def\gca{Geochim.~Cosmochim.~Acta}%
          % Geochimica Cosmochimica Acta
\def\grl{Geophys.~Res.~Lett.}%
          % Geophysics Research Letters
\def\jcp{J.~Chem.~Phys.}%
          % Journal of Chemical Physics
\def\jgr{J.~Geophys.~Res.}%
          % Journal of Geophysics Research
\def\jqsrt{J.~Quant.~Spec.~Radiat.~Transf.}%
          % Journal of Quantitiative Spectroscopy and Radiative Trasfer
\def\memsai{Mem.~Soc.~Astron.~Italiana}%
          % Mem. Societa Astronomica Italiana
\def\nphysa{Nucl.~Phys.~A}%

\def\physrep{Phys.~Rep.}%
          % Physics Reports
\def\physscr{Phys.~Scr}%
          % Physica Scripta
\def\planss{Planet.~Space~Sci.}%
          % Planetary Space Science
\def\procspie{Proc.~SPIE}%
          % Proceedings of the SPIE

\newcommand{\rmp}{Rev. Mod. Phys.}
\newcommand{\ijmpd}{Int. J. Mod. Phys. D}
\newcommand{\sovjetp}{Soviet J. Exp. Theor. Phys.}
\newcommand{\jkas}{J. Korean. Ast. Soc.}
\newcommand{\PPVI}{Protostars and Planets VI}
\newcommand{\njp}{New J. Phys.}
\newcommand{\rap}{Res. Astro. Astrophys.}

%%%%%%%%%%%%%%%%%%%%%%%%%%%%%%%%%%%%%%%%%%%%%%
%%%%%%%%%%%%%%%%%%%%%%%%%%%%%%%%%%%%%%%%%%%%%%

\setcounter{chapter}{5}

\chapter[Primordial Gas Collapse In The Presence of Radiation: Direct Collapse Black Hole or Population III star]{Primordial Gas Collapse in The Presence of Radiation:\\ Direct Collapse Black Hole or Population III star?}\footnotetext{$^1$ Preprint~of~a~review volume chapter to be published in Latif, M., \& Schleicher, D.R.G., ''Primordial Gas Collapse in The Presence of Radiation:\\ Direct Collapse Black Hole or Population III star?'', Formation of the First Black Holes, 2018 \textcopyright Copyright World Scientific Publishing Company, https://www.worldscientific.com/worldscibooks/10.1142/10652}\label{BA_chapter}

\author[Bhaskar Agarwal]{Bhaskar Agarwal}

\address{Universit{\"a}t Heidelberg, Zentrum f{\"u}r Astronomie, Institut f{\"u}r Theoretische Astrophysik, Albert-Ueberle-Str. 2, D-69120 Heidelberg, Germany, \\ bhaskar.agarwal@uni-heidelberg.de }

\begin{abstract}
The first billion years in the evolution of the Universe mark the formation of the first stars, black holes and galaxies. The radiation from the first galaxies plays an important role in determining the final state of primordial gas collapsing in a neighboring halo. This is due to the fact that the primary coolant for primordial gas is molecular hydrogen, which can be dissociated into atomic hydrogen by Lyman-Werner photons in the energy range $11.2 - 13.6$~eV. While cooling by molecular hydrogen leads to Pop. III star formation, cooling by atomic hydrogen can lead to the formation of a supermassive star (or quasi-star) which results in the formation of a massive $10^{4-5} \Msun$ black hole, or a direct collapse black hole. The spectrum of this radiation field is critical in order to determine whether a primordial gas cloud forms a Pop. III star or a  very massive black hole. We will in the following explore this scenario and discuss how the radiation spectrum influences the outcome of the collapse.

\end{abstract}

\body

\setcounter{page}{115}

\section{Introduction}
\subsection{First Stars or Direct Collapse Black Holes }
%- {cooling curves, minihalo vs. atomic cooling halo, Pop. III in atomic cooling halo}\\

\begin{figure}
\centering
\includegraphics[angle=90,width=0.9\columnwidth,trim = 0.7cm 0.2cm 0.7cm .5cm, clip]{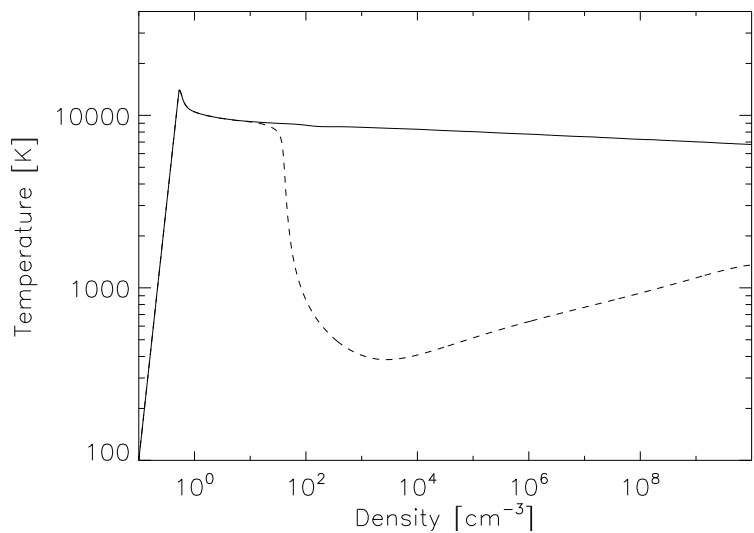}
\caption{Density-temperature diagram for primordial gas collapse. Cooling via atomic H leads to a DCBH (solid), while cooling via molecular hydrogen leads to Pop.~III stars (dashed). Figure taken from \citet{Agarwal2016}, reproduced by permission of Oxford University Press / on behalf of the RAS.}
\label{fig.rho-T}
\end{figure}

As discussed in Chapter 3, the primary coolant in the early Universe is H$_2$, which can cool the gas down to $\sim 200$~K. Cooling of primordial gas by molecular hydrogen leads to Pop. III star formation (see Chapter 4), thus we must understand the formation and destruction of H$_2$ in the context of primordial star formation. Partial dissociation of molecular hydrogen can delay Pop. III star formation in a minihalo, as the halo needs to grow and regain its H$_2$ fraction to efficiently cool and collapse into stars. As discussed in Chapter 5, in the complete absence of H$_2$, atomic hydrogen assumes the role of the primary coolant which can cool the gas to $\sim 8000$~K. Comparing the Jeans mass, M$_J$, at a number density of $n=10^3\rm \ cm^{-3}$, H$_2$ cooling at 200 K leads to a M$_J \sim 283 \Msun$, while atomic hydrogen (H) cooling at 8000 K leads to a M$_J = 10^5\Msun$. The Jeans mass resulting from H cooling is readily available in atomic cooling haloes, i.e. haloes with T$_{vir} \sim 10^4$~K, and thus the idea of supermassive stars that undergo runaway collapse into black holes (BHs) was born \citep{Rees78,Omukai2001}. The black holes formed through this scenario are commonly referred to as direct collapse black holes (DCBH). To outline the basic thermodynamical framework, the density-temperature digram for gas undergoing collapse and cooling via H$_2$ (leading to Pop. III stars, dashed line) vs. cooling via H (leading to DCBH, solid line) is shown in Fig.~\ref{fig.rho-T}. Thus, any channel that leads to the formation or destruction of H$_2$ becomes critically important in the gas collapse process. 

\begin{table}[htp]
\caption{default}
\begin{center}
\begin{tabular}{|c|c|}
Abbreviation & full name\\ \hline
BH & black hole\\
DCBH & direct collapse black hole\\
LW & Lyman-Werner\\
SED & spectral energy distribution\\
SFR & star formation rate\\
SF & star formation
\end{tabular}
\end{center}
\label{abrsuper}
\end{table}%

\subsection{Soft-Ultra Violet radiation feedback}
%- {moderate LW radiation field}

As soon as the first population of stars appears in the Universe, subsequently leading to the formation of the second generation of stars, the effects of stellar radiation become relevant for primordial gas collapse. Lyman-Werner (LW) photons in the energy range 11.2 - 13.6 eV can photo-dissociate H$_2$ into H via the reaction

\begin{equation}
\rm H_2 + h\nu \rightarrow \rm H + H.
\end{equation}\\
Thus LW radiation from the first stars (and galaxies) can dissociate H$_2$ and delay Pop. III star formation in primordial minihaloes \citep{hrl96,Ciardi2000}. This LW radiation is often parametrised in the literature in the form of a specific intensity, J$_{\rm LW}$, in units of $10^{-21}\ \rm erg/s/cm^2/Hz/sr$. In order to form Pop. III stars a primordial minihalo in the presence of LW radiation, the mass of the halo needs to be above a critical threshold \citep{machacek01,OShea2008}, M$_{th}$, or in other words

\begin{equation}
\rm M_{th} \sim 4\left( 1.25\times10^5 + 8.7\times 10^5 (4\pi J_{LW})^{0.47}\right) \Msun.
\end{equation}\\
Setting J$_{\rm LW} = 1$ in the above equation sets the threshold mass to $10^7\Msun$ or the atomic cooling limit. Thus, one might argue from that DCBH formation should ensue in a minihalo exposed to J$_{\rm LW} > 1$.
However, detailed studies \citep{Shang2010,Latif2014} show that in the presence of a LW background, only
the molecular hydrogen in the outer regions of minihaloes is dissociated
and a considerable fraction of molecular hydrogen (0.001) still
exists in the central region of the haloes. In order to completely suppress Pop. III star formation, it is essential to lower the H$_2$ fraction
 to $10^{-8}$ which can be achieved by extremely high values of LW specific intensity as discussed in Sec.~\ref{sec.BB}.

\subsection{Importance of H$^-$ photo-detachment}
\label{sec.h_minus}

While H$_2$ is an obvious determinant of the fate of primordial gas collapse, another important molecular species is that of H$^-$. At low densities an important pathway for the formation of H$_2$ is via H$^-$ where
\begin{equation}
\label{eq.form_Hm}
\rm H + e \rightarrow H^- + e,
\end{equation}
\begin{equation}
\label{eq.form_H2}
\rm H^- + H \rightarrow H_2 + e. 
\end{equation}\\

Thus it is important to understand the role of H$^-$ in pristine gas collapse as well. A stellar radiation field can help in destroying H$^-$ via photodetachment by photons in the energy range $h\nu \geq 0.754~$eV,

\begin{equation}
\rm H^- + h\nu \rightarrow \rm H + e.
\end{equation}

Due to the importance of this species, it has only recently been argued \citep{WolcottGreen2012} that besides UV, the infrared (IR) component of radiation is also critical to primordial gas collapse. This highlights the role of low mass stars in the early Universe which are the primary producers of IR photons. We refer here to Chapter 3 for a detailed discussion on the chemical processes regarding H$^-$.

\section{Modeling gas collapse under the influence\\ of a stellar radiation field}
%- kde, kdi \\
%- introduce the Jc paramter. \\
%- parameter that determines whether youre making H2 or destroying it \\

\noindent The photodestruction of the two most relevant species, i.e. photodetachment of H$^-$, and the photodestruction of H$_ 2$ is parametrised as

\begin{eqnarray}
\rm k_{di} = C_{di}\alpha J_{LW} \ \rm s^{-1},\\
\rm k_{de} = C_{de}\beta J_{ LW}\ \rm s^{-1},
\end{eqnarray}\\
where C$_{di} = 1.38\times10^{-12}$ and C$_{ de} = 1.1\times 10^{-10}$ are rate constants, $\alpha$ and $\beta$ are rate parameters that depend on the spectral shape of the irradiating source, and J$_ {LW}$ is the specific intensity  in the LW as defined earlier. In order to better understand the rate parameters, we first define the quantity L$_n$ with the spectral energy distribution (SED) normalised to be 1 at $h\nu = 13.6$~eV, in units of $K_{21}$. Then, $\alpha$ is defined as 
\begin{equation}
\alpha = \frac{\int\limits_{\nu_{0.76}}^{\nu_{13.6}} \frac{4\pi {\rm L}_{n}}{h\nu}\sigma_{\nu} d\nu}{1.1\times10^{-10}}, 
\label{eq.alpha}
\end{equation}

where ${\nu_{0.76}}\ \&\ {\nu_{13.6}}$ correspond to the frequency limits of 0.76 and 13.6 eV, respectively. The cross section denoted by $\sigma_{\nu}$ is written as

\begin{equation}
\sigma_{\nu} = 10^{-18}\lambda^3(\frac{1}{\lambda} - \frac{1}{\lambda_o})^{1.5}\sum\limits_{n=1}^{6}\rm C_n[\frac{1}{\lambda} - \frac{1}{\lambda_o}]^{\frac{n-1}{2}}\ cm^2,
\end{equation}

where $\lambda_o\approx1.6 \mu m$ (0.76eV) and the wavelength is in $\mu m$. Lastly, the parameter $\rm C_n$ is of a tabulated form \citep{John1988,deJong1972,Wishart1979,Tegmark1997,Abel97}.

The parameter $\beta$ on the other hand is defined with a much simpler form

\begin{equation}
\beta =\rm \frac{1}{\Delta \nu_{\rm LW}}\int\limits^{\nu_{13.6}}_{\nu_{11.2}}L_{n}d\nu, 
\label{eq.beta}
\end{equation}
 
where ${\nu_{11.2}}\ \&\ {\nu_{13.6}}$ correspond to the frequency limits 11.2 and 13.6 eV respectively (i.e. the LW band), and $\Delta \nu_{LW} = {\nu_{13.6}} - {\nu_{11.2}}$. 

It is now clear that for a given level of flux (or specific intensity) the assumed spectral shape of the irradiating source, that determines the spectral parameters, can affect the rates of destruction of H$_ 2$ and H$^-$. 

%The parameter that captures the competition between the formation and desctruction of H$_2$ or the effective formation rate k$_{\rm form}$ \citep{Omukai:2001p128} is defined as: 
%
%\begin{equation}
%{\rm k_{form}} \equiv {\rm k_\ref{eq.form_Hm}} \frac{{\rm k_\ref{eq.form_H2}} n_{\rm{H}}}{n_{\rm{H}} {\rm k_\ref{eq.form_H2}}+ {\rm k_{\rm{de}}}}
%\label{eq.netform}
%\end{equation}
%
%where the hydrogen number density is denoted by $n_H$, and the indices of the individual rate coefficients k$_i$ refer to the equation number of the reaction discussed previously in this chapter. 

\subsection{First galaxies as single temperature black bodies}
\label{sec.BB}
%- Om01, Shang+10
%- First estimates of Jc 

The first attempts \citep{Omukai2001} to model stellar radiation attributed a single temperature black body (BB), or a power-law spectrum to the first galactic populations. A Pop. III stellar cluster was modeled using a T$=10^5$~K (T5) black body, and Pop. II stellar clusters with a T$=10^4$~K black body (T4). The presence of such an external radiation field led to the conclusion that primordial gas can withstand collapse into Pop. III type protostellar cores for a high level of incident flux. In the absence of H$_2$, where molecular hydrogen has been dissociated into H due to the incident radiation, the gas can only cool down to $\sim 8000$~K, thereby yielding a Jeans mass of $\sim 10^6\Msun$ at $n\sim10^3\ \rm cm^{-3}$. Such high gas masses are readily available in atomic cooling haloes with a virial temperature T$_{vir}\sim10^4$~K, thus giving rise to the idea of the formation of a single supermassive star (or quasi-star) that can \textit{directly} collapse into a massive $10^{4-5}\Msun$ BH. 

The first study \citep{Omukai2001} that explored the minimum value of the LW flux at which the gas undergoes isothermal collapse at 8000 K, or J$_{crit}$, was done using a 1D model, that set the flux for a power law type spectrum to J$_{crit} = 10^5$ and a T4 spectrum to J$_{crit} = 10^3$. Exposing the gas to fluxes higher than J$_{crit}$ leads to a similar chemo-thermodynamical state as seen at J$_{crit}$. A decade later, another study \citep{Shang2010} revisited the idea and using their 3D gas collapse simulations found a significantly lower J$_{crit} \sim 30-100$ for a T4 type spectrum, while for a T5 spectrum they found $J_{crit} = 10^{4-5}$. Besides an updated methodology, they attributed this difference to the softer shape of the T4 spectrum, which leads to a much more effective destruction of H$^-$ at lower densities.

In other words, one can understand this difference in terms of the reaction rates, k$_{de}$ and k$_{di}$ defined earlier. Using Eqs. \ref{eq.alpha} and \ref{eq.beta}, for a T4 type spectrum we obtain $\alpha \sim 2000,\ \beta \sim0.9$. Thus for a J$_{crit} = 30$ the rates are evaluated as

\begin{eqnarray}
& \rm k_{de} = &1.1\times10^{-10} \times 2000 \times 30 = 6\times10^{-6}\ \rm s^{-1},\\
& \rm k_{di} =  &1.38 \times10^{-12} \times 0.9 \times 30 = 3.3\times10^{-11}\ \rm s^{-1}.
\end{eqnarray}\

Similarly for a T5 type spectrum, we obtain $\alpha \sim 0.1,\ \beta \sim 3$, and for a J$_{crit}=10^4$ the rates are

\begin{eqnarray}
& \rm k_{de} = &1.1\times10^{-10} \times 0.1 \times 10^4 = 1\times10^{-7}\ \rm s^{-1},\\
& \rm k_{di} =  &1.38 \times10^{-12} \times 3 \times 10^4 = 4.14\times10^{-8}\ \rm s^{-1}.
\end{eqnarray}\

When compared with the typical formation rate of H$_2$ at a density of $n = 10^3\rm cm^{-3}$, both the above pairs of values for k$_{de}$ and k$_{di}$ are able to dissociate enough molecular hydrogen into atomic hydrogen leading to isothermal collapse of the pristine gas \citep{Shang2010}. Note that the values of k$_{de}$ for both the T4 and T5 spectra are similar $\sim 10^{-7} \ \rm s^{-1}$, which can be understood in terms of the interplay between the rate parameter $\alpha$ and J$_{crit}$ for the respective spectra. The H$_2$ photodissociation rate however varies by 3 orders of magnitude between the two spectra. This requires investigation of the relative importance of k$_{de}$ and k$_{di}$ for a given value of J$_{LW}$.

\subsection{First galaxies from realistic spectra}
%- A14: alpha and beta vary a lot \\
%- Sugimura: ratio of rates \\
%- 3D collapse of clouds: Latif et al., Regan et al. not completely consistent but deviation from old Jc (hints at a new Jc)\\
%- A16, WG16: no single Jc\\
%- push that the J depends on where you intersect the curve\\

\begin{figure}
\centering
\includegraphics[width=0.9\columnwidth,trim = 0.7cm 0.2cm 0.7cm 12.25cm, clip]{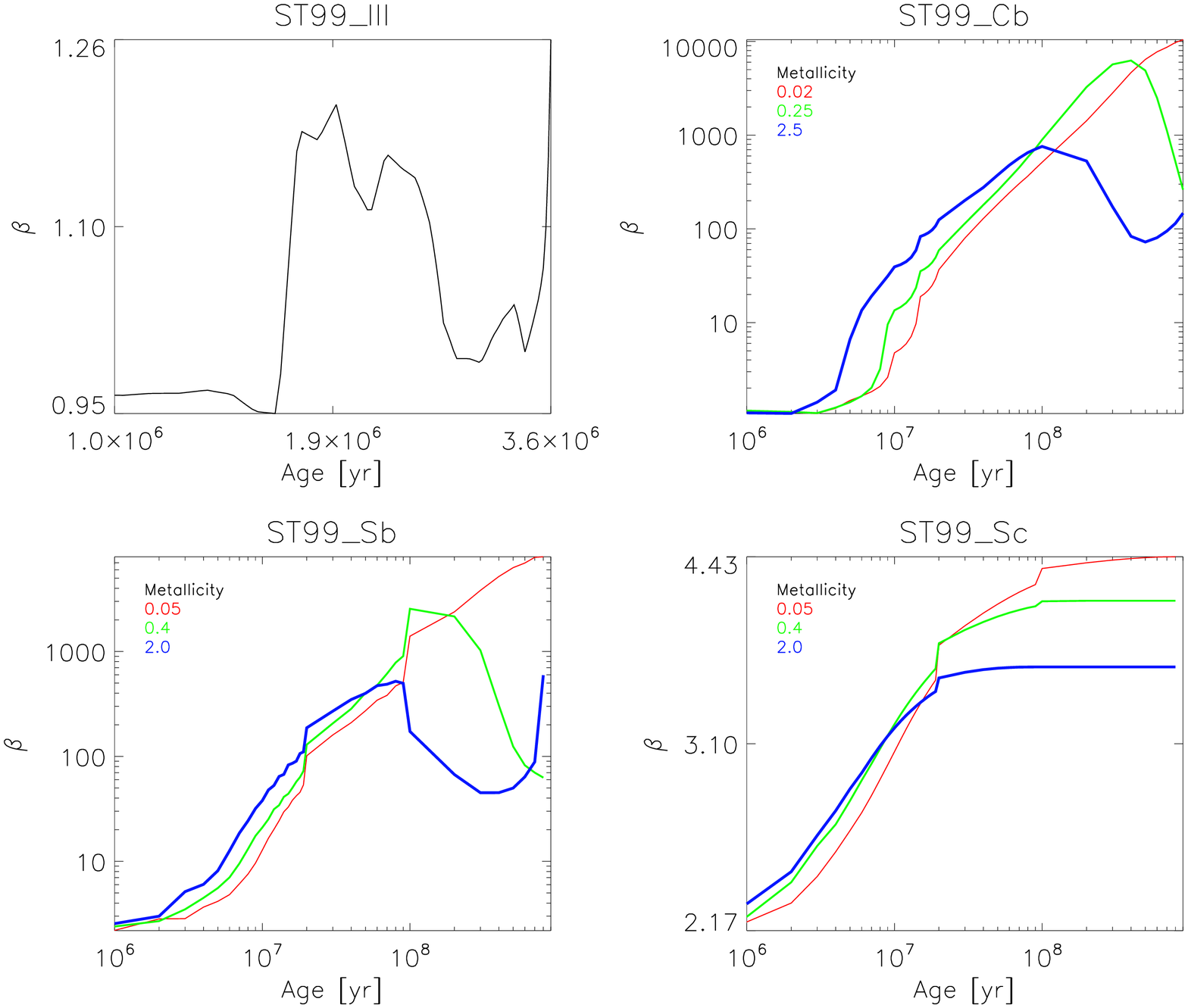}
\includegraphics[width=0.9\columnwidth,trim = 0.7cm 0.2cm 0.7cm 12.25cm, clip]{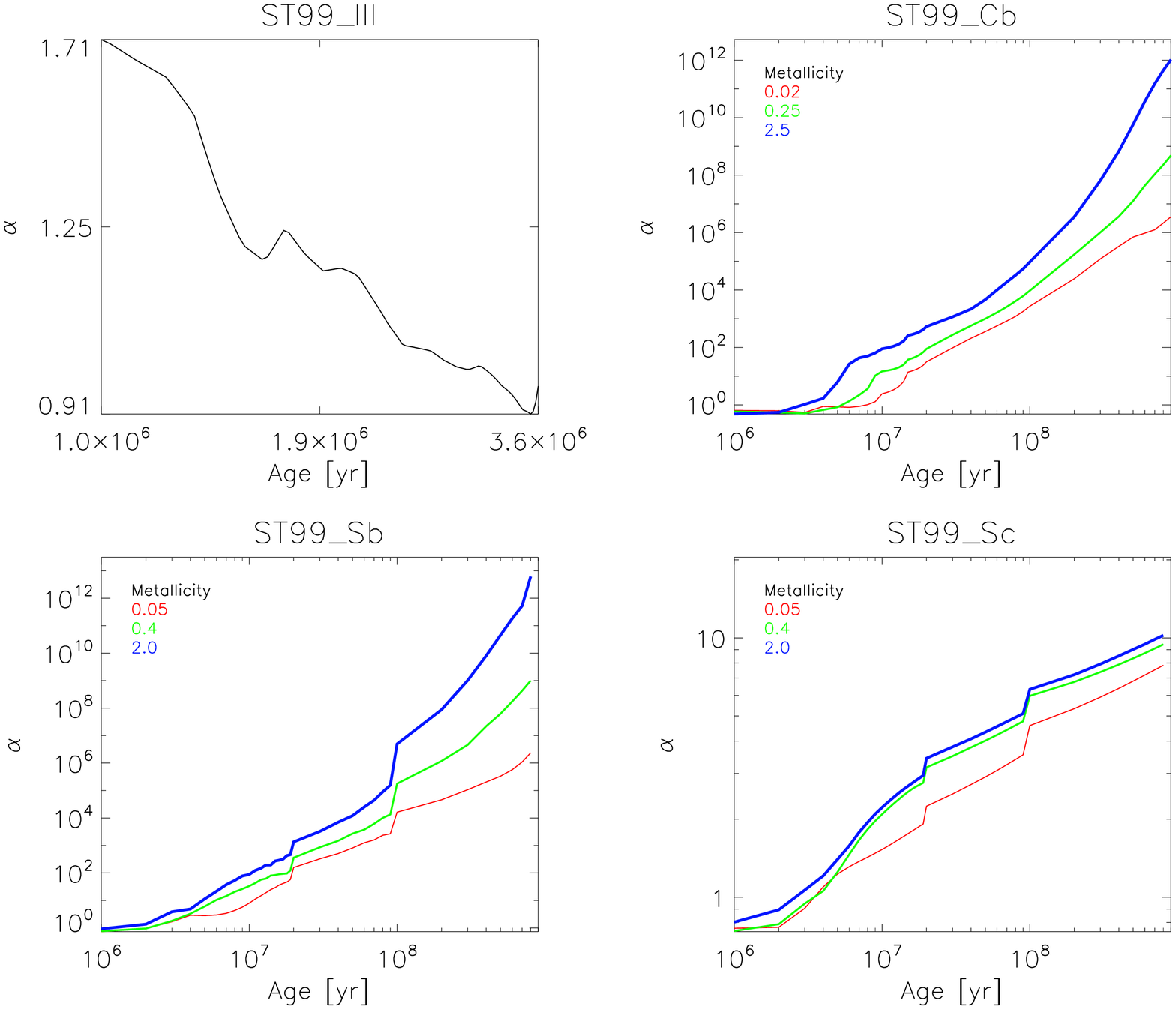}
\caption[Reaction rate coefficients for realistic spectra]{The rate parameters for stellar populations modeled as {\sc starburst99} SEDs. The metallicity is quoted in solar units with ${Z}_{\odot} =0.02$. In the left panels a single burst of $10^6\Msun$ is used to model the SFR, whereas in right panels, a continuous star formation rate of $1\Msun \rm / yr$ is employed (Figure reproduced \citet{Agarwal2015} by permission of Oxford University Press / on behalf of the RAS). }
\label{fig.alpha-beta-A15}
\end{figure}

Attributing a single blackbody to either a Pop. III or Pop. II type stellar population is a strong approximation. One can better understand the behaviour of J$_{crit}$ for \textit{realistic} galaxy samples by first studying the values of the rate parameters.
Attributing single stellar populations, extracted from {\sc Starburst99} \citep{Leitherer1999}, to the first galaxies (Pop. II type), one can obtain the values of $\alpha$ and $\beta$ for a range of stellar parameters such as the stellar mass (M$_{\star}$), star formation rate (SFR), metallicity (Z), and age (t$_{\star}$). For SEDs computed from {\sc starburst99} (or from \citet{Agarwal_binary} binary stellar synthesis models) using a Salpeter IMF, the rate parameters are plotted in Fig.~\ref{fig.alpha-beta-A15}, where star formation (SF) is modeled either as a single burst of 10$^6\,\Msun$ that is then allowed to age, or in the form of a continuous SFR of $1\Msun/\rm yr$. The rate parameters vary over several orders of magnitude for the burst mode, and less than an order of magnitude for a continuous mode of SF \citep{Agarwal2015}. This suggests that approximating stellar populations as single temperature black bodies where the rate parameters are constant for any given star formation history is an inaccurate way of capturing the effects of their radiation on pristine gas collapse.

It is now understood that the radiation constraint for isothermal collapse of pristine gas into a BH can not be understood as a single value of J$_{crit}$, but rather in terms of the interplay between the relevant photo-destruction rates. %In other words one can understand the radiation requirement for DCBH in either of the following two ways
%The first attempt to capture this interplay was in the form of the ratio of k$_{de}$ and k$_{di}$ \cite{Sugimura:2014p3946}. One can associate an effective blackbody temperature to a given value of $k_{de} / k_{di}$, and this formulation is particularly useful if one already knows the rates. However note that this computation was made using a fixed k$_{di}$ as the LW band is quite narrow for the spectral shape to play a role. An elegant fit using the parameter $x = log_{10}(k_{de}/k_{di}) -2$ was thus proposed:
%
%\begin{eqnarray}
%\rm if\ x\geq 0,& \ \ & \rm J_{cr}  =  1400\\
%\rm if\ x < 0,&  \ \ & \rm J_{cr}  =   1400 \times 10^{(a_1x + a_2x^2)}
%\end{eqnarray}
%
%where $a_1 = -0.19$ and $a_2= -0.12$.\\

\begin{figure}
\centering
\includegraphics[width=0.9\columnwidth]{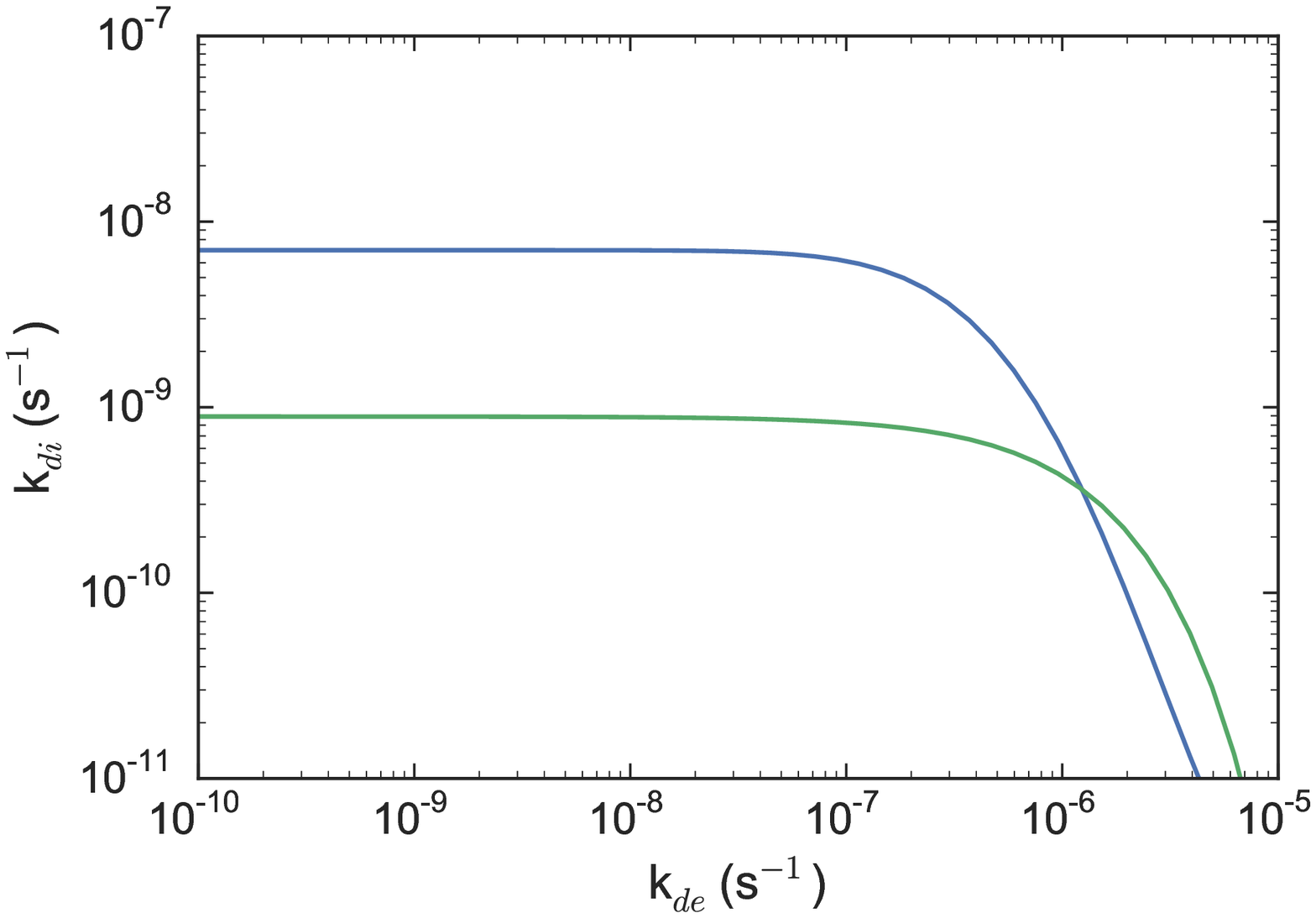}
\caption[Critical curve for direct collapse]{The critical curve in the k$_{de}$ - k$_{di}$ phase space above which the gas undergoes isothermal collapse into a DCBH, and below which the gas cools by H$_2$ leading to the formation of Pop. III stars. The difference between the curves is mostly due to the difference in the jeans length and self-shielding of H$_2$, where the blue curve \citep{Agarwal2016} is obtained with a jeans length that is twice that of the one used in the green curve. Figure adopted from \citet{Wolcott2017}, reproduced by permission of Oxford University Press / on behalf of the RAS.}
\label{fig.sugimura}
\end{figure}

A more robust formulation of the critical LW radiation field required for DCBH formation comes in the form of a critical curve in the k$_{de}$ and k$_{di}$ phase space \citep{Agarwal2016,Wolcott2017}. One can draw a curve in the k$_{de}$ - k$_{di}$ parameter space to differentiate the region of DCBH formation from that of fragmentation into Pop. III stars. The curve represents the minimum value of k$_{di}$ needed for a given value of k$_{de}$ at which the gas collapses isothermally. The curve itself does not change for a given irradiating source-DCBH halo arrangement and can be viewed as a fundamental relation. However,  an irradiating galaxy can occupy any part of this phase space depending on its spectral shape and separation from the target halo (or the site of DCBH), thereby determining to the chemo-thermodynamical  evolution of the pristine gas in the target halo.

\section{Other radiation effects}

%- Ionising radiation\\
%- X- rays\\
%- escape fractions\\
%\section{Appendices}\index{appendix}
%If there is more than one appendix, number them alphabetically.
%Sectional units are obtained with the

\subsection{The Escape Fraction of LW photons \citep{Kitayama2004,Schauer2015,Schauer17}}
Although the effects of radiation on gas collapse are broadly understood, the escape of radiation from the host haloes of stellar components is still under investigation. The escape fraction of LW photons from the first generation of stars forming in massive haloes is close to zero, and varies from $0 - 85\%$ in smaller haloes, depending on the stellar mass. Higher mass Pop. III stars ($>10 \Msun$) in low mass haloes ($\sim 10^{5-6}\Msun$) generally lead to larger escape fractions of up to $\sim 80\%$. On the other hand, escape fractions of LW photons from the first galaxies in the Universe, primarily composed of the second generation or Population II stars, are generally higher than $60\%$. Thus, the stellar populations of the first galaxies can assume their role as the main contributors to the LW radiation field in the early Universe.

\subsection{Effects of H ionising radiation \citep{Johnson2014,Chon2017}}
Photons in the energy range $h\nu > 13.6\ \rm eV$ are capable of ionising atomic H and can thus heat up the gas in the first haloes, preventing subsequent collapse. If the gas in the haloes is self-shielded to ioninsing photons, the effects of ionising radiation are minimal and DCBH formation in atomic cooling haloes can proceed as usual. However if the gas densities are low (e.g. in the early stages of a halo's evolution), such that self-shielding is not efficient, ionising photons can delay collapse of gas in atomic cooling haloes by $\sim 20-30$~Myr. Furthermore, if the collapsing halo is close enough to the radiation source the gas in it can even get photo-evaporated, thus halting the collapse entirely. However, these calculations are heavily dependent on the modelling of the propagation of the ionising photons, or in other words the inter-galactic-medium (i.e. gas between the ionising source and the collapsing halo) and the inter-stellar-medium (i.e. gas within the ionising source).

\subsection{Effects of X-rays \citep{Inayoshi2011,Latif2015,Inayoshi2015,gl16,Regan2016}}

X-rays can prove to be critical to the scenario of DCBH formation as they can penetrate dense gas and ionise H, thereby increasing the fraction of free electrons which can lead to the formation of H$_2$ via the reactions discussed in Sec.~\ref{sec.h_minus}. Thus, presence of X-ray photons can thus lead to the J$_{crit}$ that is higher than the original one, i.e. in the absence of X-ray photons. The X-ray specific intensity is defined as J$_{x}$ in the same units as J$_{\rm LW}$ which facilitates an easy comparison and investigation of the relative importance of the two quantities. 
The current consensus on the topic is that for values of J$_{x} < 0.01$, the revised J$_{crit}$ needed for DCBH formation is at most a factor of two higher than the original one. The J$_{crit}$ needed for larger values of J$_{x}$ is even higher and can be upto 10 times higher than the original one if J$_{x} \sim 1$. The field still lacks a full investigation of this topic in the form of detailed 3D hydrodynamical simulations, and the results discussed here still employ BB type spectra to represent stellar populations. 

With the presented chapter, we conclude the discussion regarding the uncertainties due to microphysical processes in black hole formation scenarios. In the following Chapter 7, we will investigate the role of collisional processes in the formation of massive black holes, while the evolution of supermassive stars is described in Chapter 8. Growth and feedback, statistical predictions and a comparison with current and future observations are given in subsequent chapters.

%%%%%%%%%%%%%%%%%%%%%%%%%%%%%%%%%%%%%%%%%%%%%%
%%%%%%%%%%%%%%%%%%%%%%%%%%%%%%%%%%%%%%%%%%%%%%

{
\bibliographystyle{ws-rv-har}    % author-date citation/references %%% WS 04-Dec-17
\bibliography{ref}
}

\printindex[aindx]           % to print author index
\printindex                  % to print subject index

\end{document}